\documentclass[twocolumn,preprintnumbers,amsmath,amssymb,superscriptaddress]{revtex4-2}
\UseRawInputEncoding

\usepackage{latexsym,amssymb,amsthm,amsmath,epsfig,braket}
\usepackage[left=2.00cm, right=2.00cm, top=2.00cm, bottom=2.00cm]{geometry}
\usepackage{xcolor}
\usepackage{svg}
\usepackage[hidelinks]{hyperref}
\usepackage{float}
\usepackage{soul}

\usepackage{bm}
\begin{document}
\title{Floquet Engineering of Topological Phases and Magneto-Optical Response in a Driven $d$-wave Altermagnet}
% Force line breaks with \\
\author{Muzami Shah}
\email{muzamil@qau.edu.pk}
\affiliation{Department of Physics, Quaid-i-Azam University, Islamabad 45320, Pakistan}

%\date{\today}% It is always \today, today, %  but any date may be explicitly specified

\begin{abstract}
We study how Floquet driving with linearly polarized light controls the topology and magneto-optical response of a two-dimensional (2D) $d$-wave altermagnet. In the absence of linearly polarized optical field and under spin conservation, we find that the system hosts a spin-Chern (a quantum-spin-Hall analog) phase with Chern numbers of opposite sign in the two spin sectors. The irradiated optical field breaks the $C_{4z}\mathcal{T}$ crystalline antiunitary symmetry between the spin sectors. Symmetry breaking originates from polarization-dependent Peierls phases, which renormalize hopping anisotropically along the two axes. The resulting spin-selective gap closures produce intermediate Chern-insulating phases with $C=\pm1$. The drive amplitude $A_0$ determines the inversion thresholds, while rotating the polarization by $\pi/2$ swaps the spin sectors and reverses the Chern number. Using the Kubo formalism, we compute the frequency-dependent longitudinal and Hall conductivities and derive the corresponding Faraday and Kerr rotations for a free-standing conducting sheet. The longitudinal response tracks the Floquet-renormalized interband thresholds, whereas the optical Hall response, together with the sign of the magneto-optical rotations, distinguishes the two opposite Berry-curvature chiralities. Sizable Kerr angles occur only within narrow resonant windows and should be interpreted together with the reflected intensity and Kerr ellipticity. These results identify linearly polarized light as a symmetry-selective handle for spin-resolved band inversion, Chern-number switching, and contact-free optical detection in $d$-wave altermagnets.

\end{abstract}

\maketitle
\section{Introduction}
Altermagnetism introduces an exceptional class of magnetic behavior by combining features traditionally associated with both ferromagnets and antiferromagnets \cite{Smejkal2022a,Smejkal2022c,Yuan2020,Ahn2019}. It originates from the interplay between non-relativistic magnetic order and crystal rotational symmetries, generating momentum-dependent spin splitting while maintaining a vanishing net magnetization in real space \cite{Mazin2021,Bai2023}. In particular, $d$-wave altermagnets can generate spin-polarized currents through their spin-split electronic structure even in the absence of spin-orbit coupling \cite{GonzalezHernandez2021,Bose2022}. In these systems, time-reversal symmetry $\mathcal{T}$ is broken, whereas a combined symmetry involving time reversal and a nontrivial crystal point-group operation remains preserved. In addition to momentum-resolved spin polarization, altermagnets can exhibit physical phenomena that are generally absent in conventional collinear antiferromagnets, including the anomalous Hall effect, the multipiezoelectric effect, and controllable spin polarization \cite{Leeb2024,Krempasky2024,Bai2023,Zhang2024}. Their intrinsically spin-split band structures also provide a promising platform for realizing nontrivial topological phases \cite{Li2025,Rao2024,Ma2024,Antonenko2025,Fernandes2024,Li2024}. 

The electronic structure of matter can be strongly modified by external stimuli. Light is particularly attractive in this context because it offers ultrafast and noncontact control of intrinsic electronic properties while providing access to nonequilibrium states that are absent in the static regime. Periodic light irradiation, commonly described within Floquet theory, has emerged as a powerful approach for dynamically engineering electronic structures and topological phases \cite{Dubey2025FloquetTransport}. Floquet engineering has enabled the realization of a broad range of nonequilibrium phenomena \cite{Oka2019FloquetEngineering,Rudner2020BandStructure}, including light-induced Hall effects \cite{Oka2009PhotovoltaicHall,Kitagawa2011TransportProperties,McIver2019LightInduced,Sato2019MicroscopicTheory}, Floquet topological insulators \cite{Kitagawa2010TopologicalCharacterization,Lindner2011FloquetTopological}, light-induced Weyl semimetals \cite{Fu2017PhaseTransitions}, and Floquet time crystals \cite{Liu2024HigherOrder}.

More recently, the optical manipulation of electronic, magnetic, and topological properties in altermagnets has received increasing attention \cite{liu2026ultrafast,zou2025floquet}. Proposed light-induced phenomena include odd-parity magnetism \cite{Liu2026LightInducedOddParityAltermagnets,Li2025StackingSliding,7bss-9yxb}, dynamically generated higher-order spin-orbit coupling \cite{Ghorashi2025DynamicalGeneration,fu2026floquet,xt23-9pnv}, circularly polarized light-induced quantum anomalous Hall phases \cite{Zou2025FloquetQAH,k3xb-8pts}, and other forms of optically tunable topology \cite{Ganguli2025TunableTopology}. Previous studies of light-induced anomalous Hall effects and Chern-insulating phases have predominantly relied on circularly polarized light, which explicitly breaks time-reversal symmetry $\mathcal{T}$ and can also break parity-time-reversal symmetry $\mathcal{PT}$ \cite{Oka2009PhotovoltaicHall,Kitagawa2011TransportProperties,Xu2021LightInducedQAH,Wang2018LightInducedTypeII,Zhu2023FloquetEngineering}. By contrast, linearly polarized light generally preserves $\mathcal{T}$ and is therefore usually ineffective in generating anomalous Hall responses in nonmagnetic materials and conventional antiferromagnets.

% PROBLEM IN ORIGINAL LINE 101: conventional spin Chern terminology was used although
% ordinary time reversal is broken by the altermagnetic order.

Altermagnets offer a distinct setting in this respect: their magnetic order already breaks conventional time-reversal symmetry $\mathcal{T}$, yet the two spin blocks can still be related by a crystalline antiunitary operation such as $C_{4z}\mathcal{T}$. In the spin-conserving square-lattice model, this operation interchanges $k_x\leftrightarrow k_y$ and $\uparrow\leftrightarrow\downarrow$, forcing the charge Hall responses to cancel in the undriven spin-Chern (spin Chern-like) phase. A fixed linearly polarized pump is generally not invariant under $C_{4z}$ rotation, since it renormalizes the $x$- and $y$-directed hopping channels by different Bessel factors. The pump thus breaks $C_{4z}\mathcal{T}$ through this polarization-selective hopping anisotropy, allowing a finite anomalous Hall response to emerge without optical helicity, an external magnetic field, or reversal of the magnetic order parameter.

Magneto-optical phenomena serve as sensitive probes of magnetic order, band topology, and transverse optical response in quantum materials~\cite{Sato2022Fundamentals}. Faraday rotation captures the polarization rotation of transmitted light, while Kerr rotation is extracted from the reflected field~\cite{Sato2022Fundamentals,Tse2011MagnetoOptical}. Such probes are especially valuable in compensated magnets, where vanishing net magnetization can coexist with spin-split bands and a finite optical Hall conductivity~\cite{Rao2024Tunable}. Earlier studies of altermagnetic magneto-optics have focused mainly on equilibrium spin-momentum locking, optical selection rules, Rashba coupling, external fields, or static strain~\cite{Liu2026MOKE,Rao2024Tunable,Sun2025SymmetryBreaking}. In contrast, the control parameter considered here is the polarization direction of a linearly polarized Floquet pump. Its anisotropic renormalization of the hopping amplitudes breaks $C_{4z}\mathcal{T}$, splits the spin-resolved inversion thresholds, and yields Hall and magneto-optical responses that are switchable via the pump polarization.

%The two spin sectors undergo band-gap closings and reopenings at different driving amplitudes due to the resulting symmetry breaking. This results in a spin-dependent redistribution of the Berry curvature and also enables transitions from a quantum spin Hall phase to spin-polarized Chern-insulating phases with total Chern number $C=\pm1$, followed by a transition to a topologically trivial insulating phase. The corresponding light-induced anomalous Hall conductivity is directly reflected in the frequency-dependent Faraday and Kerr rotations. It provides experimentally accessible optical signatures of the topological transitions. Therefore, the central novelty of this work is the identification of linearly polarized light as a symmetry-selective dynamical control knob for simultaneously changing crystalline symmetry, band topology, anomalous Hall transport, and magneto-optical response in a two-dimensional altermagnet. This mechanism does not require circularly polarized light, an external magnetic field, static Rashba coupling, or a finite equilibrium magnetization.

We employ a four-band tight-binding model on a two-dimensional square lattice and derive its leading-order, off-resonant time-averaged Floquet Hamiltonian. From this effective Hamiltonian we obtain the spin-resolved band gaps, Berry curvature, Fukui--Hatsugai--Suzuki Chern numbers, optical conductivities, and magneto-optical rotations. The drive amplitude sets when each spin block closes and reopens its gap, while the polarization direction determines which spin block inverts first. As a result, the two-dimensional $(A_0,\theta)$ parameter space contains a spin-Chern phase, two oppositely signed Chern phases with $C=-1$ and $C=+1$, and a topologically trivial phase.

Within linear response, we evaluate the longitudinal and Hall optical conductivities of the weak probe using the Kubo formalism. The linearly polarized pump drives an anisotropic state, so the Faraday and Kerr rotations are extracted from the full conductivity tensor using matrix transmission and reflection coefficients; the usual circular-channel formulas follow only in the isotropic limit. The sign of the Hall response identifies the two opposite Chern chiralities, whereas its resonant strength is set by probe frequency, broadening, and Floquet-renormalized transition energies. We therefore emphasize polarization-induced sign changes and their correspondence with phase boundaries rather than the peak rotation angle itself.

\section{Model and Theoretical Formalism}
\label{sec:model}
We consider a four-band two-dimensional out-of-plane $d$-wave altermagnetic
model on a square lattice~\cite{Liu2026LPLAM,Ma2024AMTI,GonzalezHernandez2025SpinChern,Ayesha2026}. The basis is chosen as
\begin{equation}
\left\{
p_z\uparrow,\,
\frac{1}{\sqrt{2}}\left(d_{xz}\uparrow+i d_{yz}\uparrow\right),\,
p_z\downarrow,\,
\frac{1}{\sqrt{2}}\left(d_{xz}\downarrow-i d_{yz}\downarrow\right)
\right\}.
\label{eq:basis}
\end{equation}
In this basis, the Hamiltonian is block diagonal in the spin degree of freedom,
\begin{equation}
H(\bm{k})
=
\begin{pmatrix}
H_{\uparrow}(\bm{k}) & 0\\
0 & H_{\downarrow}(\bm{k})
\end{pmatrix},
\label{eq:H0}
\end{equation}
where each spin block is a two-band Hamiltonian,
\begin{equation}
H_{\sigma}(\bm{k})
=
\bm{d}^{\sigma}(\bm{k})\cdot\bm{\tau}
=
\sum_{\alpha=x,y,z}
d_{\alpha}^{\sigma}(\bm{k})\tau_{\alpha},
\qquad
\sigma=\uparrow,\downarrow .
\label{eq:Hsigma_static}
\end{equation}
Here $\bm{\tau}=(\tau_x,\tau_y,\tau_z)$ are Pauli matrices acting in the
orbital subspace. For the spin-up block, the static $d$-vector components are
\begin{align}
d_x^{\uparrow}(\bm{k})
&=
-v\sin k_x,
\nonumber\\
d_y^{\uparrow}(\bm{k})
&=
-vt_a\sin k_y,
\nonumber\\
d_z^{\uparrow}(\bm{k})
&=
m+b\left(\cos k_x+t_a^2\cos k_y\right).
\label{eq:d_static_up}
\end{align}
The spin-down block is obtained by exchanging $k_x$ and $k_y$:
\begin{equation}
H_{\downarrow}(k_x,k_y)
=
H_{\uparrow}(k_y,k_x).
\label{eq:spin_down_relation}
\end{equation}
Therefore,
\begin{align}
d_x^{\downarrow}(\bm{k})
&=
-v\sin k_y,
\nonumber\\
d_y^{\downarrow}(\bm{k})
&=
-vt_a\sin k_x,
\nonumber\\
d_z^{\downarrow}(\bm{k})
&=
m+b\left(\cos k_y+t_a^2\cos k_x\right).
\label{eq:d_static_down}
\end{align}
Here $m$ is the on-site potential, $v$ is the inter-orbital hopping amplitude,
$b$ controls the band-inversion term, and $t_a$ is the anisotropy parameter.
When $t_a=1$, the system restores $\mathcal{PT}$ symmetry and reduces to a
conventional antiferromagnet. When $t_a\neq1$, the anisotropy between the
$x$ and $y$ directions produces the $d$-wave altermagnetic spin splitting.
The spin-up and spin-down sectors are related by crystalline-time-reversal
symmetries such as $C_{4z}\mathcal{T}$ or the mirror symmetry
$\mathcal{M}_{xy}$.

The corresponding band energies are
\begin{equation}
\varepsilon_s^{\sigma}(\bm{k})
=
s\left|
\bm{d}^{\sigma}(\bm{k})
\right|,
\qquad
s=\pm ,
\label{eq:static_eigenvalues}
\end{equation}
where $s=+$ denotes the conduction band and $s=-$ denotes the valence band. In the spin-conserving model, the band-inverted regime
\begin{equation}
|m|<|b|\left(1+t_a^2\right)
\label{eq:spin Chern_condition}
\end{equation}
has opposite spin-block Chern numbers and is therefore a spin-Chern (spin Chern-like) insulator. %Because ordinary $\mathcal{T}$ is broken by the altermagnetic order, this terminology should not be confused with a conventional $\mathbb{Z}_2$ quantum spin Hall insulator protected solely by time reversal. The robustness of the counterpropagating edge modes against spin-mixing perturbations depends on the crystalline symmetry and should be tested separately.

\section{Floquet Effective Hamiltonian under Linearly Polarized Light}
\label{sec:floquet}

Under periodic driving, $H(\bm{k},t)=H(\bm{k},t+T)$ with $T=2\pi/\omega$, the Floquet modes satisfy
\begin{equation}
[H(\bm{k},t)-i\partial_t]u_\lambda(\bm{k},t)=\epsilon_\lambda(\bm{k})u_\lambda(\bm{k},t),
\end{equation}
with $u_\lambda(\bm{k},t+T)=u_\lambda(\bm{k},t)$. Expanding
\begin{equation}
H(\bm{k},t)=\sum_\ell H_\ell(\bm{k})e^{i\ell\omega t},
\end{equation}
the off-resonant high-frequency expansion gives, to leading order,
\begin{equation}
H_{\rm eff}(\bm{k})=H_0(\bm{k})
+\sum_{\ell\geq1}\frac{[H_\ell(\bm{k}),H_{-\ell}(\bm{k})]}{\ell\omega}
+\mathcal{O}(\omega^{-2}).
\label{eq:high_frequency_expansion}
\end{equation}

We take a linearly polarized vector potential
\begin{equation}
\bm{A}(t)=A_0(\cos\theta\cos\omega t,\,\sin\theta\cos\omega t)
\end{equation}
(with $e/\hbar$ absorbed into $A_0$) and apply the Peierls substitution $\bm{k}\to\bm{k}+\bm{A}(t)$. Since $\bm{A}(t)=\bm{A}(-t+\tau)$, the commutator term in Eq.~\eqref{eq:high_frequency_expansion} vanishes, and the leading correction is the time-averaged Hamiltonian. The Jacobi--Anger expansion renormalizes the bond hoppings by zeroth-order Bessel functions,
\begin{align}
t_x^{\rm eff}&=t_x\,J_0(A_0a\cos\theta)\equiv t_xj_1,\\
t_y^{\rm eff}&=t_y\,J_0(A_0a\sin\theta)\equiv t_yj_2.
\label{eq:directional_hopping}
\end{align}
Because $\bm{A}(t)$ projects unequally onto the two bond directions, $j_1\neq j_2$ in general, and the driven lattice acquires optical anisotropy even though the bare hopping is isotropic (equality holds only for $A_0=0$ or $|\cos\theta|=|\sin\theta|$). This treatment requires the off-resonant, high-frequency condition: the pump photon energy $\hbar\Omega_{\rm p}$ must exceed the band gap and the electronic bandwidth. The dimensionless amplitude is
\begin{equation}
A_0=\frac{eaE_0}{\hbar\Omega_{\rm p}};
\end{equation}
converting to physical field/intensity units requires specifying $a$, $v$, the bandwidth, and $\Omega_{\rm p}$, together with a check that real absorption remains weak.

$H_{\rm eff}(\bm{k})$ remains block diagonal, $H_{\rm eff}^\sigma(\bm{k})=\bm{d}_{\rm eff}^\sigma(\bm{k})\cdot\bm{\tau}$, with
\begin{align}
d_{x,\rm eff}^{\uparrow}&=-vj_1\sin k_x,
\label{eq:dx_up}\\
d_{y,\rm eff}^{\uparrow}&=-vt_aj_2\sin k_y,
\label{eq:dy_up}\\
d_{z,\rm eff}^{\uparrow}&=m+bj_1\cos k_x
\nonumber\\
&\quad+bj_2t_a^2\cos k_y,
\label{eq:dz_up}
\end{align}
and
\begin{align}
d_{x,\rm eff}^{\downarrow}&=-vj_2\sin k_y,
\label{eq:dx_down}\\
d_{y,\rm eff}^{\downarrow}&=-vt_aj_1\sin k_x,
\label{eq:dy_down}\\
d_{z,\rm eff}^{\downarrow}&=m+bj_2\cos k_y
\nonumber\\
&\quad+bj_1t_a^2\cos k_x.
\label{eq:dz_down}
\end{align}
The bands are
\begin{equation}
\varepsilon_s^\sigma(\bm{k})=s\left|\bm{d}_{\rm eff}^\sigma(\bm{k})\right|,\qquad s=\pm,
\end{equation}
giving a spin-resolved direct gap
\begin{equation}
\Delta_\sigma(\bm{k})=2\left|\bm{d}_{\rm eff}^\sigma(\bm{k})\right|.
\label{eq:direct_gap}
\end{equation}

Since altermagnetic order already breaks conventional time-reversal symmetry, the relevant relation between the two spin blocks is the crystalline antiunitary operation $C_{4z}\mathcal{T}$, which exchanges $k_x\leftrightarrow k_y$ and $\uparrow\leftrightarrow\downarrow$. Linearly polarized light carries no optical helicity, so it does not induce helicity-driven time-reversal breaking as circularly polarized light would. However, a fixed polarization axis is not $C_{4z}$-invariant:
\begin{equation}
C_{4z}\mathcal{T}\,H_{\rm eff}(k_x,k_y;\theta)\,(C_{4z}\mathcal{T})^{-1}
\nonumber
\end{equation}
\begin{equation}
=H_{\rm eff}(k_x,k_y;\theta+\pi/2),
\end{equation}
which differs from $H_{\rm eff}(k_x,k_y;\theta)$ whenever $j_1\neq j_2$, so the pump breaks $C_{4z}\mathcal{T}$ via directional hopping anisotropy. A $\pi/2$ polarization rotation swaps $j_1\leftrightarrow j_2$ and the two spin blocks, reversing which spin mass inverts first: the surviving topological spin-up block gives $C=-1$, while the rotated configuration makes the spin-down block topological with $C=+1$. The Chern number changes only via a bulk gap closing and reopening.

\subsection{Chern number calculation}

The driven spin sectors have topological character, which is obtained from the occupied-band Berry curvature of the two-band Hamiltonian given in Eq.~\eqref{eq:Heff_spin}. For each spin block $\sigma=\uparrow,\downarrow$, the effective terms can be written as
\begin{equation}
H_{\rm eff}^{\sigma}(\bm{k})=\bm{d}_{\rm eff}^{\sigma}(\bm{k})\cdot\bm{\tau},
\qquad
\hat{\bm d}^{\sigma}=\frac{\bm d_{\rm eff}^{\sigma}}{|\bm d_{\rm eff}^{\sigma}|}.
\label{eq:Heff_spin}
\end{equation}
The Berry curvature of the occupied band becomes
\begin{equation}
\Omega_-^{\sigma}(\bm{k})
=-\frac{1}{2}\hat{\bm d}^{\sigma}\cdot
\left(
\frac{\partial \hat{\bm d}^{\sigma}}{\partial k_x}
\times
\frac{\partial \hat{\bm d}^{\sigma}}{\partial k_y}
\right).
\end{equation}
The spin-resolved Chern number is obtained from berry curvatures as
\begin{equation}
C_{\sigma}=\frac{1}{2\pi}\int_{\rm BZ}\Omega_-^{\sigma}(\bm{k})\,d^2k,
\end{equation}
and the total and spin Chern numbers are calculated as
\begin{equation}
C=C_{\uparrow}+C_{\downarrow},\qquad
C_s=\frac{C_{\uparrow}-C_{\downarrow}}{2}.
\end{equation}
The Chern number can be evaluated numerically using the gauge-invariant Fukui-Hatsugai-Suzuki lattice method. 
%The link variables on a discretized Brillouin zone are given as
%\begin{equation}
%U_x(\bm{k})=
%\frac{\langle u_-(\bm{k})|u_-(\bm{k}+\Delta k_x)\rangle}
%{|\langle u_-(\bm{k})|u_-(\bm{k}+\Delta k_x)\rangle|},
%\end{equation}
%\begin{equation}
%U_y(\bm{k})=
%\frac{\langle u_-(\bm{k})|u_-(\bm{k}+\Delta k_y)\rangle}
%{|\langle u_-(\bm{k})|u_-(\bm{k}+\Delta k_y)\rangle|}.\end{equation}
%The lattice Berry flux through each plaquette is
%\begin{equation}
%F_{xy}(\bm{k})={\rm Im}\ln
%\left[
%U_x(\bm{k})U_y(\bm{k}+\Delta k_x)
%U_x^{-1}(\bm{k}+\Delta k_y)U_y^{-1}(\bm{k})
%\right],
%\end{equation}
%and the Chern number are
%\begin{equation}
%C_{\sigma}=\frac{1}{2\pi}\sum_{\bm{k}}F_{xy}^{\sigma}(\bm{k}).
%\end{equation}
%This method is stable because it does not require a globally smooth gauge for the eigenvectors. The Chern number changes only when the bulk gap closes and reopens, so the phase boundaries in the $(A_0,\theta)$ plane correspond to values where $|\bm d_{\rm eff}^{\sigma}(\bm{k})|=0$ for one spin sector. Away from the gap-closing boundaries, the spin-resolved Chern number $C_{\sigma}$ remains strictly quantized, providing an internal numerical consistency check for the phase diagram. In this regime, the topological classification is governed by the global Berry-curvature integral rather than by the instantaneous magnitude of the band gap.

\section{Optical Conductivity from the Kubo Formula}
\label{sec:optical_conductivity}
We calculate the frequency-dependent optical conductivity from the same
Floquet effective Hamiltonian in Eq.~\eqref{eq:Heff_spin}. The spin-resolved
velocity operator is defined by
\begin{equation}
\hat v_i^\sigma(\bm{k})
=
\frac{1}{\hbar}
\frac{\partial H_{\rm eff}^{\sigma}(\bm{k})}{\partial k_i},
\qquad
i=x,y .
\label{eq:velocity_definition}
\end{equation}
Using Eq.~\eqref{eq:Heff_spin}, this becomes
\begin{equation}
\hat v_i^\sigma(\bm{k})
=
\frac{1}{\hbar}
\sum_{\alpha=x,y,z}
\frac{\partial d_{\alpha,\rm eff}^{\sigma}(\bm{k})}{\partial k_i}
\tau_{\alpha}.
\label{eq:velocity_dvector}
\end{equation}

For the spin-up sector, the explicit velocity operators are
\begin{align}
\hat v_x^\uparrow(\bm{k})
&=
\frac{1}{\hbar}
\left[
-vj_1\cos k_x\,\tau_x
-bj_1\sin k_x\,\tau_z
\right],
\label{eq:vx_up}
\\
\hat v_y^\uparrow(\bm{k})
&=
\frac{1}{\hbar}
\left[
-vt_a j_2\cos k_y\,\tau_y
-bj_2t_a^2\sin k_y\,\tau_z
\right].
\label{eq:vy_up}
\end{align}
For the spin-down sector, one obtains
\begin{align}
\hat v_x^\downarrow(\bm{k})
&=
\frac{1}{\hbar}
\left[
-vt_a j_1\cos k_x\,\tau_y
-bj_1t_a^2\sin k_x\,\tau_z
\right],
\label{eq:vx_down}
\\
\hat v_y^\downarrow(\bm{k})
&=
\frac{1}{\hbar}
\left[
-vj_2\cos k_y\,\tau_x
-bj_2\sin k_y\,\tau_z
\right].
\label{eq:vy_down}
\end{align}

Following the Kubo formalism, the spin-resolved optical conductivity tensor
for finite photon momentum $\bm{Q}$ is
\begin{widetext}
\begin{equation}
\begin{split}
\sigma_{ij}^{\sigma}(\bm{Q},\omega)
=
-i e^2\hbar
\int_{\rm BZ}
\frac{d^2k}{(2\pi)^2}
\sum_{s,s'=\pm}
&
\frac{
f\!\left[\varepsilon_s^\sigma(\bm{k})\right]
-
f\!\left[\varepsilon_{s'}^\sigma(\bm{k}+\bm{Q})\right]
}{
\varepsilon_s^\sigma(\bm{k})
-
\varepsilon_{s'}^\sigma(\bm{k}+\bm{Q})
}
\\
&\times
\frac{
\left\langle
u_s^\sigma(\bm{k})
\left|
\hat v_i^\sigma(\bm{k})
\right|
u_{s'}^\sigma(\bm{k}+\bm{Q})
\right\rangle
\left\langle
u_{s'}^\sigma(\bm{k}+\bm{Q})
\left|
\hat v_j^\sigma(\bm{k})
\right|
u_s^\sigma(\bm{k})
\right\rangle
}{
\varepsilon_s^\sigma(\bm{k})
-
\varepsilon_{s'}^\sigma(\bm{k}+\bm{Q})
+
\hbar\omega
+
i\hbar\gamma
}.
\end{split}
\label{eq:kubo_finiteQ}
\end{equation}
\end{widetext}
Here $i,j=x,y$, $\gamma$ is a phenomenological broadening parameter, and
$f[\varepsilon_s^\sigma(\bm{k})]$ is the Fermi--Dirac distribution function.
The photon momentum is negligible in the optical limit, so we take
\begin{equation}
\bm{Q}\rightarrow0 .
\label{eq:long_wavelength}
\end{equation}
Then Eq.~\eqref{eq:kubo_finiteQ} reduces to
\begin{widetext}
\begin{equation}
\begin{split}
\sigma_{ij}^{\sigma}(\omega)
=
-i e^2\hbar
\int_{\rm BZ}
\frac{d^2k}{(2\pi)^2}
\sum_{s,s'=\pm}
&
\frac{
f\!\left[\varepsilon_s^\sigma(\bm{k})\right]
-
f\!\left[\varepsilon_{s'}^\sigma(\bm{k})\right]
}{
\varepsilon_s^\sigma(\bm{k})
-
\varepsilon_{s'}^\sigma(\bm{k})
}
\\
&\times
\frac{
\left\langle
u_s^\sigma(\bm{k})
\left|
\hat v_i^\sigma(\bm{k})
\right|
u_{s'}^\sigma(\bm{k})
\right\rangle
\left\langle
u_{s'}^\sigma(\bm{k})
\left|
\hat v_j^\sigma(\bm{k})
\right|
u_s^\sigma(\bm{k})
\right\rangle
}{
\varepsilon_s^\sigma(\bm{k})
-
\varepsilon_{s'}^\sigma(\bm{k})
+
\hbar\omega
+
i\hbar\gamma
}.
\end{split}
\label{eq:kubo_long}
\end{equation}
\end{widetext}

The total optical conductivity is obtained by summing the two spin-sector
contributions:
\begin{equation}
\sigma_{ij}(\omega)
=
\sigma_{ij}^{\uparrow}(\omega)
+
\sigma_{ij}^{\downarrow}(\omega).
\label{eq:spin_sum_sigma}
\end{equation}
In particular, the longitudinal optical conductivity is
\begin{equation}
\sigma_{xx}(\omega)
=
\sigma_{xx}^{\uparrow}(\omega)
+
\sigma_{xx}^{\downarrow}(\omega),
\label{eq:sigma_xx}
\end{equation}
and the optical Hall conductivity is
\begin{equation}
\sigma_{xy}(\omega)
=
\sigma_{xy}^{\uparrow}(\omega)
+
\sigma_{xy}^{\downarrow}(\omega).
\label{eq:sigma_xy}
\end{equation}

For a low-temperature insulating state with the Fermi level inside the gap,
the valence band is occupied and the conduction band is empty:
\begin{equation}
f\!\left[\varepsilon_-^\sigma(\bm{k})\right]=1,
\qquad
f\!\left[\varepsilon_+^\sigma(\bm{k})\right]=0 .
\label{eq:occupation}
\end{equation}
Thus, the dominant optical process is the interband transition
$\varepsilon_-^\sigma(\bm{k})\rightarrow\varepsilon_+^\sigma(\bm{k})$.
Using the direct transition energy in Eq.~\eqref{eq:direct_gap}, the
interband optical conductivity can be written as
\begin{widetext}
\begin{equation}
\begin{split}
\sigma_{ij}^{\sigma}(\omega)
=
i e^2\hbar
\int_{\rm BZ}
\frac{d^2k}{(2\pi)^2}
\frac{
M_{ij}^{\sigma}(\bm{k})
}{
\Delta_{\sigma}(\bm{k})
}
\left[
\frac{1}{
\hbar\omega-\Delta_{\sigma}(\bm{k})+i\hbar\gamma
}
-
\frac{1}{
\hbar\omega+\Delta_{\sigma}(\bm{k})+i\hbar\gamma
}
\right],
\end{split}
\label{eq:interband_optical}
\end{equation}
\end{widetext}
where
\begin{equation}
M_{ij}^{\sigma}(\bm{k})
=
\left\langle
u_-^\sigma(\bm{k})
\left|
\hat v_i^\sigma(\bm{k})
\right|
u_+^\sigma(\bm{k})
\right\rangle
\left\langle
u_+^\sigma(\bm{k})
\left|
\hat v_j^\sigma(\bm{k})
\right|
u_-^\sigma(\bm{k})
\right\rangle .
\label{eq:Mij}
\end{equation}

For the longitudinal optical response,
\begin{equation}
M_{xx}^{\sigma}(\bm{k})
=
\left|
\left\langle
u_-^\sigma(\bm{k})
\left|
\hat v_x^\sigma(\bm{k})
\right|
u_+^\sigma(\bm{k})
\right\rangle
\right|^2 .
\label{eq:Mxx}
\end{equation}
For the optical Hall response,
\begin{equation}
M_{xy}^{\sigma}(\bm{k})
=
\left\langle
u_-^\sigma(\bm{k})
\left|
\hat v_x^\sigma(\bm{k})
\right|
u_+^\sigma(\bm{k})
\right\rangle
\left\langle
u_+^\sigma(\bm{k})
\left|
\hat v_y^\sigma(\bm{k})
\right|
u_-^\sigma(\bm{k})
\right\rangle .
\label{eq:Mxy}
\end{equation}

Equation~\eqref{eq:Mxx} provides that the longitudinal optical conductivity is controlled by the absolute transition strength $|\langle u_-^{\sigma}|\hat v_x^{\sigma}|u_+^{\sigma}\rangle|^2$. Therefore, the strength of allowed interband transitions driven by an electric field along the $x$ direction is measured by $\sigma_{xx}$. On the other hand, Eq.~\eqref{eq:Mxy} has the product of two velocity matrix elements in orthogonal directions. The Hall response depends on the relative phase between the $x$- and $y$-direction of optical transition amplitudes and this is why it is sensitive to the chirality of the Bloch wave functions and the Berry curvature. The longitudinal conductivity shifts with $A_0$, and the reason for this is contained in the Floquet-renormalized transition energy. Since LPL changes the Hamiltonian through the Bessel factors $j_1=J_0(A_0\cos\theta)$ and $j_2=J_0(A_0\sin\theta)$, the transition energy becomes $\Delta_{\sigma}(\bm{k};A_0,\theta)=2|\bm d_{\rm eff}^{\sigma}(\bm{k};A_0,\theta)|$. If the Floquet-renormalized gap decreases, the absorption peak redshifts to lower photon energy; if the gap increases, the peak blueshifts. Thus, $\sigma_{xx}$ shifts because LPL modifies the direct interband transition energies.

The sign reversal of $\sigma_{xy}$ follows from the reversal of Berry-curvature chirality at a gap-closing transition. Near a closing point, the Hamiltonian reduces to a massive anisotropic Dirac model,
\begin{equation}
\Omega(\bm{q})\propto
-\frac{m_{\rm eff}v_xv_y}
{(v_x^2q_x^2+v_y^2q_y^2+m_{\rm eff}^2)^{3/2}}.
\end{equation}
When the Floquet drive reverses $m_{\rm eff}$, the Berry curvature and the low-frequency Hall response reverse sign. For a gapped phase with the chemical potential in the gap, the clean dc limit provides the essential normalization check,
\begin{equation}
\lim_{\omega\rightarrow0}\operatorname{Re}\sigma_{xy}(\omega)
=C\frac{e^2}{h}
=\frac{C}{2\pi}\frac{e^2}{\hbar}.
\label{eq:dcHallcheck}
\end{equation}
To check the numerical normalization, we recalculated the low-frequency Hall conductivity while reducing the broadening and refining the momentum grid. In the gapped $C=\pm1$ phases, $\operatorname{Re}\sigma_{xy}(\omega\rightarrow0)$ converges to $\pm1/(2\pi)$ in units of $e^2/\hbar$, whereas it vanishes in the spin Chern-like and trivial phases. The Chern numbers from the Fukui--Hatsugai--Suzuki method remain quantized under the same refinement.

Thus, $\sigma_{xx}$ primarily identifies transition thresholds and spectral weight, whereas the sign and dc limit of $\sigma_{xy}$ connect the optical response to the Chern topology.

\begin{figure*}[ht!]
  \centering
  \includegraphics[width=1.0\linewidth]{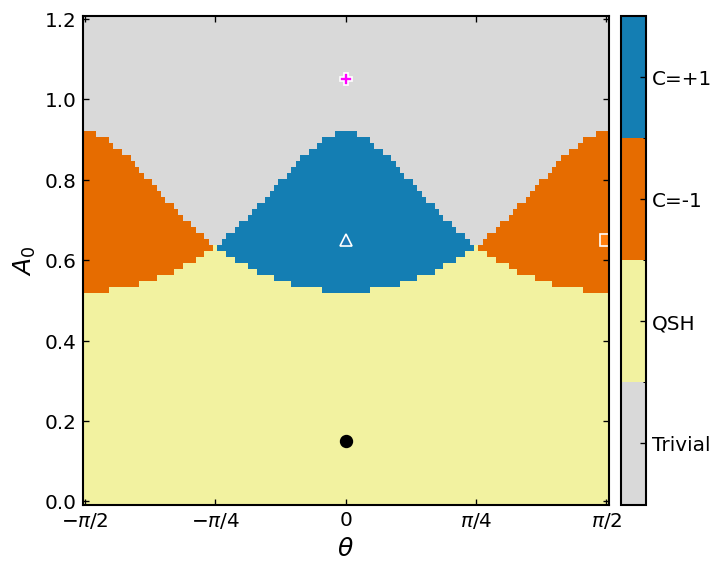}
 \caption{
Topological phase diagram of the altermagnetic spin-Chern system under linearly polarized light as a function of the drive amplitude \(A_{0}\) and polarization angle \(\theta\). The yellow, blue, purple, and gray regions denote the spin-Chern phase, the \(C=-1\) Chern-insulating phase, the \(C=+1\) Chern-insulating phase, and the trivial phase, respectively. The boundaries are obtained from spin-resolved bulk-gap closings. The black markers identify the representative parameter points used for the band-structure, Berry-curvature, conductivity, and magneto-optical calculations. Parameters: \(m=3.8v\), \(b=-v\), and \(t_a=\sqrt{3}\).}
  \label{phase}
\end{figure*}

\begin{figure*}[ht!]
  \centering
  \includegraphics[width=1.00\linewidth]{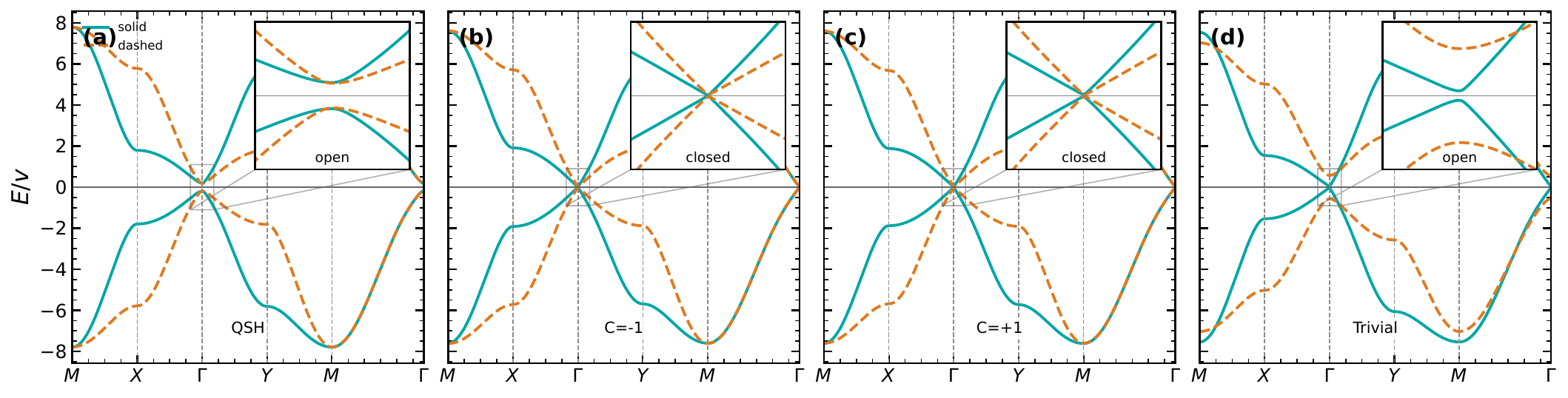}
  \caption{Spin-resolved band structures of the driven $d$-wave altermagnet for representative points in the two-dimensional $(A_0,\theta)$ phase diagram: (a) spin Chern-like phase at $(A_0,\theta)=(0.1500,0)$, (b) $C=-1$ phase at $(A_0,\theta)=(0.6250,0.8290)$, (c) $C=+1$ phase at $(A_0,\theta)=(0.6250,0.7418)$, and (d) trivial phase at $(A_0,\theta)=(1.0500,0)$. The parameters are $m=3.8v$, $b=-v$, and $t_a=\sqrt{3}$. Solid teal and dashed orange curves denote the spin-up and spin-down bands, respectively. The insets magnify the spin-dependent direct gaps around the minimum-gap region. Panels (b) and (c) correspond to points on opposite sides of the polarization-controlled boundary rather than to a common fixed-$\theta$ amplitude sweep. Rotating the pump polarization therefore switches the spin block that remains inverted and reverses the total Chern number.}
  \label{bands}
\end{figure*}
\section{Faraday rotation and ellipticity}
We now convert the Kubo conductivity into Faraday and Kerr signals for an infinitesimally thin, free-standing sheet at normal incidence. Because the linearly polarized pump produces $j_1\neq j_2$, the driven state is generally anisotropic and one must retain the complete sheet-conductance tensor,
\begin{equation}
\hat{\mathcal G}(\omega)=4\pi\alpha
\begin{pmatrix}
\sigma_{xx}(\omega)&\sigma_{xy}(\omega)\\
\sigma_{yx}(\omega)&\sigma_{yy}(\omega)
\end{pmatrix},
\label{eq:G_tensor_general}
\end{equation}
where the conductivities are expressed in units of $e^2/\hbar$. The tangential electric field is continuous and the magnetic-field discontinuity equals the sheet current. Solving these boundary conditions as a matrix equation gives
\begin{equation}
\hat{T}(\omega)=2\left[2\hat I+\hat{\mathcal G}(\omega)\right]^{-1},
\qquad
\hat{R}(\omega)=-\hat{\mathcal G}(\omega)
\left[2\hat I+\hat{\mathcal G}(\omega)\right]^{-1}.
\label{eq:JonesTR}
\end{equation}
For an incident probe polarized along $x$, $\mathbf e_i=(1,0)^T$, the transmitted and reflected Jones vectors are $\mathbf e_t=\hat T\mathbf e_i$ and $\mathbf e_r=\hat R\mathbf e_i$. For either output vector $\mathbf e=(E_x,E_y)^T$, we define
\begin{align}
S_0&=|E_x|^2+|E_y|^2,&
S_1&=|E_x|^2-|E_y|^2,\\
S_2&=2\operatorname{Re}(E_xE_y^*),&
S_3&=-2\operatorname{Im}(E_xE_y^*).
\end{align}
The rotation and ellipticity angles are then
\begin{equation}
\Theta=-\frac{1}{2}\operatorname{atan2}(S_2,S_1),
\qquad
\eta=\frac{1}{2}\sin^{-1}\!\left(\frac{S_3}{S_0}\right),
\label{eq:StokesAngles}
\end{equation}
with $(\Theta_F,\eta_F)$ evaluated from $\mathbf e_t$ and $(\Theta_K,\eta_K)$ from $\mathbf e_r$. The sign convention in Eq.~\eqref{eq:StokesAngles} matches the circular-channel convention in the isotropic limit $\sigma_{xx}=\sigma_{yy}$ and $\sigma_{yx}=-\sigma_{xy}$, but Eq.~\eqref{eq:JonesTR} remains valid when these simplifying relations do not hold. The measurable intensities are
\begin{equation}
\mathcal T=|E_{t,x}|^2+|E_{t,y}|^2,
\qquad
\mathcal R=|E_{r,x}|^2+|E_{r,y}|^2,
\end{equation}
\begin{equation}
\mathcal A=1-\mathcal R-\mathcal T.
\label{eq:RTA_general}
\end{equation}
A large Kerr angle near a reflection minimum can arise from a rapid phase variation of a weakly reflected Jones vector. We therefore interpret $\Theta_K$ only together with $\mathcal R$, $\eta_K$, and the independently calculated bulk topology.

\section{Results and discussions}
In Fig.~\ref{phase}, we have shown the topological phase diagram of the $d$ wave altermagnet subjected to a linearly polarized light as a function of the light strength \(A_{0}\) and polarization angle \(\theta\). The yellow region represents the spin-Chern phase, where the two spin sectors possess opposite Chern numbers,  leading to a zero total Chern number; however, the spin Chern number remains finite. As \(A_{0}\) increases, the linearly polarized periodic driving field renormalizes the spin-dependent mass gaps.  Due to the altermagnetic exchange, the two spin sectors are inequivalent, which causes their gaps to close and reopen at different values of \(A_{0}\) and \(\theta\). This yields intermediate Chern insulating phases with finite total Chern numbers, corresponding to the blue \(C=-1\) and purple \(C=+1\) regions. However, it must be remembered that the sign of the Chern number depends on the polarization angle \(\theta\). By increasing the strength of the optical field, the system enters the gray trivial region, where the band ordering becomes normal and both the total and spin Chern numbers vanish. 

After identifying the topological phase boundaries in Fig.~\ref{phase}, we now turn to the band evolution at representative points chosen from the \((\theta,A_0)\) phase diagram. Figures~\ref{bands}(a)--(d) display the spin-resolved band structures of the \(d\)-wave altermagnet subjected to linearly polarized light. The solid and dashed curves represent the spin-up and spin-down sectors, respectively. The phase diagram reveals two independent forms of optical control: the drive amplitude determines which topological sector is realized, while the polarization angle fixes the sign of the Chern response. This two parameter tunability enables reversible optical switching without the need for a static magnetic field or reversal of the magnetic order parameter. Throughout these calculations, we use \(m=3.8v\), \(b=-v\), and \(t_a=\sqrt{3}\). Figures~\ref{bands}(a)--(d) correspond to representative points in the two-dimensional \((A_0,\theta)\) phase diagram, not to a single fixed-\(\theta\) scan in drive amplitude. The spin Chern-like and trivial phases are taken at \(\theta=0\), while the two Chern phases of opposite sign are chosen on opposite sides of the polarization-controlled boundary. The respective points are \((A_0,\theta)=(0.1500,0)\), \((0.6250,0.8290)\), \((0.6250,0.7418)\), and \((1.0500,0)\) for the spin Chern-like, \(C=-1\), \(C=+1\), and trivial phases. Thus, panels (b) and (c) show that rotating the pump polarization switches the spin block that remains inverted and thereby reverses the total Chern number.

In Fig.~\ref{bands}(a), the system is in the spin-Chern phase. Both spin sectors remain inverted, and the bulk gap stays open, as shown in the inset. The spin-up and spin-down sectors carry opposite Chern numbers, \(C_{\uparrow}=-1\) and \(C_{\downarrow}=+1\). Therefore, the total Chern number vanishes, \(C=C_{\uparrow}+C_{\downarrow}=0\), while the spin Chern number \(C_s=(C_{\uparrow}-C_{\downarrow})/2\) remains finite. This confirms the helical topological insulating character of the spin-Chern phase. Figures~\ref{bands}(b) and (c) present the intermediate light-induced Chern insulating phases. In Fig.~\ref{bands}(b), we show the Chern insulator phase. The spin-down sector becomes topologically trivial, while the spin-up sector remains inverted, yielding Chern numbers \(C_{\uparrow} = -1\) and \(C_{\downarrow} = 0\). Thus, the total Chern number is \(C = -1\).  In Fig.~\ref{bands}(c), we observe the opposite scenario: the spin-up sector becomes trivial, while the spin-down sector remains inverted. In this case, \(C_{\uparrow}=0\) and \(C_{\downarrow}=+1\), resulting in \(C=+1\). The insets in Figs.~\ref{bands}(b) and \ref{bands}(c) depict the local gap evolution associated with these transitions. If we further increase the strength of the optical field, the system enters a trivial insulator phase. In Fig.~\ref{bands}(d), we plotted the trivial insulating phase. In this phase, the remaining inverted spin sector also undergoes gap closing and reopening, so that both spin sectors become topologically trivial. The spin-resolved Chern numbers are \(C_{\uparrow}=0\) and \(C_{\downarrow}=0\), giving \(C=0\) and \(C_s=0\). As shown in the inset, a finite bulk gap is still present; the band inversion is entirely absent. Consequently, the system loses its topological character and becomes an ordinary trivial insulator.

\begin{figure*}[ht]
  \centering		
\includegraphics[width=1.00\linewidth]{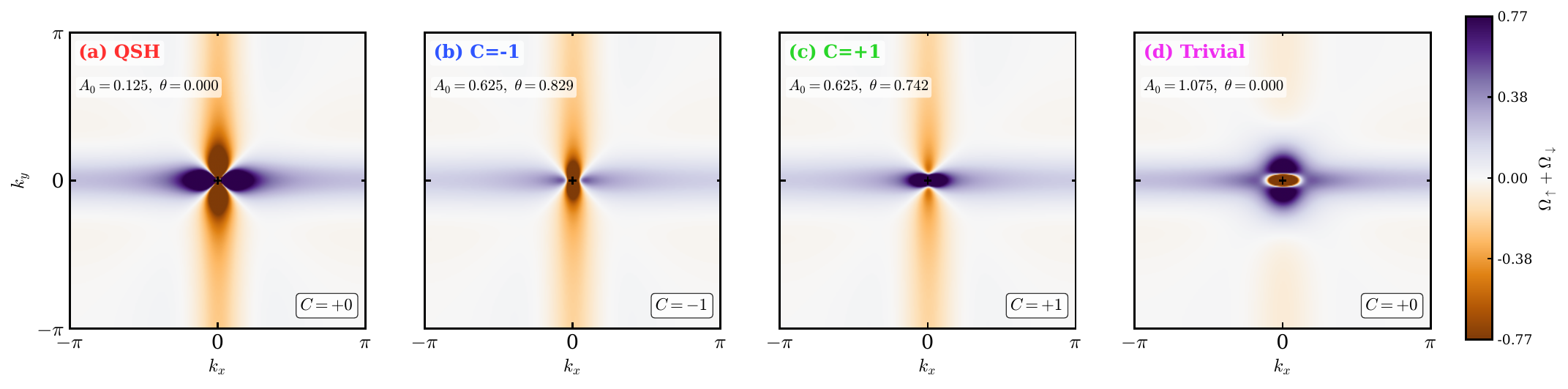}
\caption{
Berry-curvature distribution of the occupied bands in the irradiated \(d\)-wave altermagnetic spin Chern system under linearly polarized light. Panels (a)--(d) correspond to the spin Chern, \(C=-1\), \(C=+1\), and trivial phases, respectively. Red and blue regions denote Berry curvature with opposite signs. In the spin-Chern phase, the Berry curvature contributions compensate in the total charge channel, giving \(C=0\). In the intermediate Chern phases, the Berry curvature becomes sign imbalanced after a spin-selective gap closing and reopening, producing finite total Chern numbers \(C=-1\) and \(C=+1\). The reversal of the Berry-curvature pattern between panels (b) and (c) reflects the reversal of the light-induced Chern number. In the trivial phase, the local Berry curvature remains finite but integrates to zero over the Brillouin zone. The parameters are \(m=3.8v\), \(b=-v\), and \(t_a=\sqrt{3}\).
}
  \label{Berry}
\end{figure*}

Figure~\ref{Berry} illustrates the momentum-space structure of the
Berry curvature of the occupied bands across four distinct topological phases of the Floquet altermagnet. The color scale shows the local Berry curvature in the two-dimensional Brillouin zone, with red and blue regions indicating opposite signs.  In Fig.~\ref{Berry}(a), the Berry curvature in the spin-Chern phase is shown. The Berry curvature is strongly concentrated around the gap-minimum region near the Brillouin zone center. Figures~\ref{Berry}(c) and (d) correspond to two distinct, nontrivial
Floquet-Chern phases. The Berry-curvature texture in Fig.~\ref{Berry}(b) is dominated by a negative contribution around the gap-minimum region, leading to an integrated Chern number $C=-1$. Consequently, this phase possesses negative chirality and is expected to support a single chiral edge mode, with the propagation direction dictated by this topological charge. In contrast, the
$C=+1$ phase shown in Fig.~\ref{Berry}(c) exhibits the opposite Berry-curvature polarity. The transition from \(C=-1\) to \(C=+1\) signals that the Floquet field
has driven the system through a bulk-gap closing and reopening, whereby the effective mass term undergoes a sign change, and the Berry curvature reverses. Finally, Fig.~\ref{Berry}(d) shows the Berry curvature in the topologically trivial regime. The Berry curvature remains locally finite near the gap region; the positive and negative contributions compensate over the Brillouin zone, leading to a vanishing Chern number \(C=0\).  Therefore, the trivial phase does not support
a net chiral edge response. 
%Even within the trivial regime, finite local Berry curvature persists, underscoring a key distinction: optical Hall features at finite frequency can remain visible even when the net Chern number integrates to zero. A reliable topological assignment, therefore, requires the combined evidence of gap closing and reopening, full Berry curvature integration, and spin resolved Chern numbers. 

\begin{figure*}[ht!]
	\centering	
\includegraphics[width=1\linewidth]{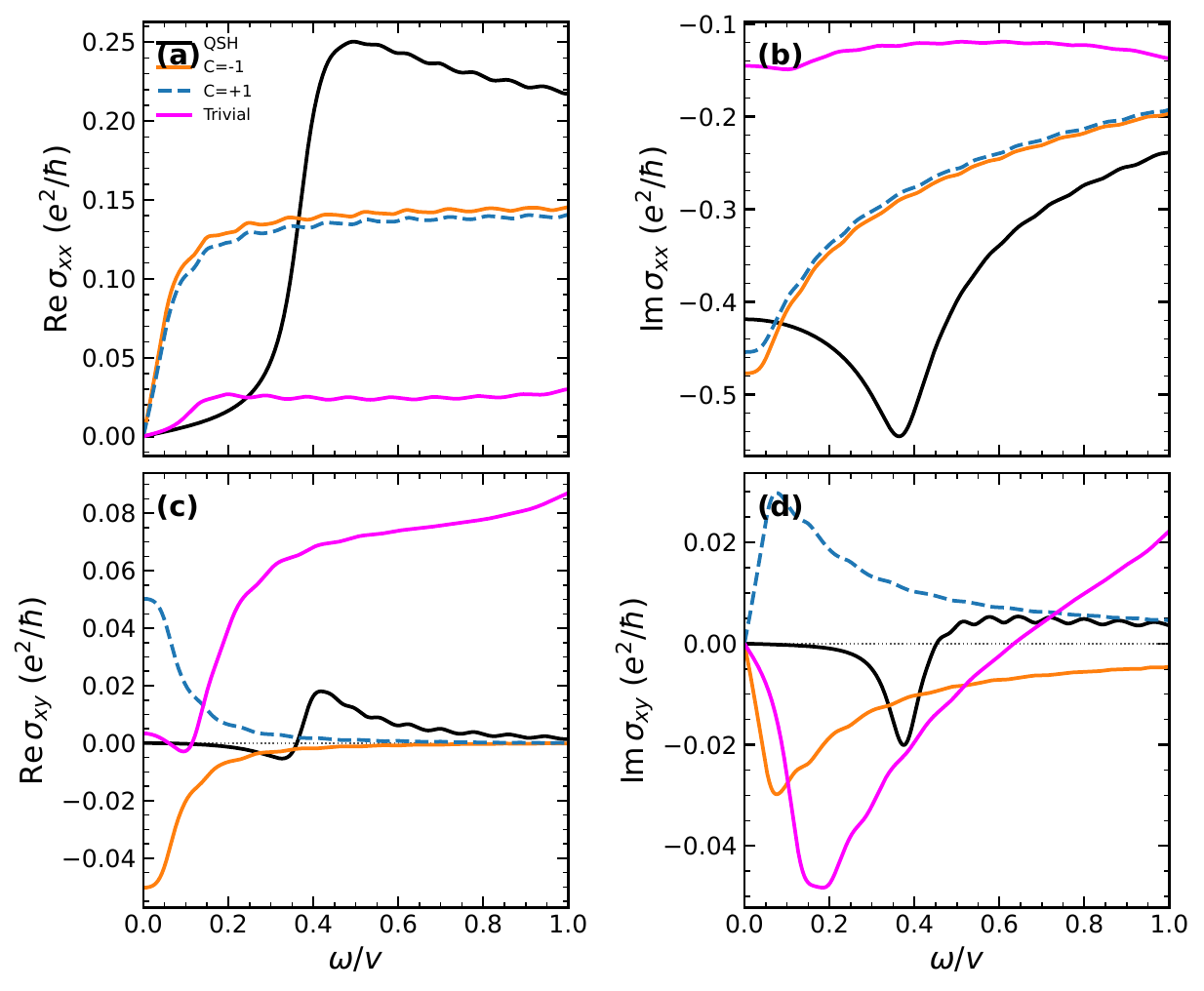}
\caption{Frequency-dependent optical conductivities in the spin Chern, $C=-1$, $C=+1$, and trivial phases: (a) $\mathrm{Re}\,\sigma_{xx}(\omega)$, (b) $\mathrm{Im}\,\sigma_{xx}(\omega)$, (c) $\mathrm{Re}\,\sigma_{xy}(\omega)$, and (d) $\mathrm{Im}\,\sigma_{xy}(\omega)$, as functions of $\omega/v$.}
\label{Conductivity}
\end{figure*}

Figure~\ref{Conductivity} shows the frequency-dependent optical response of the
Floquet-engineered altermagnet in the spin Chern, $C=-1$, $C=+1$, and trivial phases. The photon energy is plotted in units of
$\omega/v$, and the optical conductivity is expressed in units of $e^2/\hbar$. Figure~\ref{Conductivity}(a) displays the absorptive longitudinal optical conductivity,
$\mathrm{Re}\,\sigma_{xx}$ in the respective phases. At low
frequencies, $\mathrm{Re}\,\sigma_{xx}$ is strongly suppressed because
interband transitions are Pauli forbidden below the threshold energy. Whenever the incident photon energy coincides with the energy gap, resonant absorption peaks emerge in $\mathrm{Re}\,\sigma_{xx}$. A phenomenological scattering rate $\eta=0.045\,v$ is used to broaden all spectral features, giving every peak a finite width in the computed spectra.

The Chern-insulating phases (orange and blue curves in Fig.~\ref{Conductivity}(a)) display a qualitatively different optical response compared to that of the altermagnetic-spin-Chern phase. In these phases, the Floquet optical field lifts the near-degeneracy of the spin-resolved excitation channels and modifies the effective band gaps. Consequently, the resonant absorption peak occurs at a lower photon energy than in the altermagnetic-spin-Chern phase. Both the $C=-1$ and $C=+1$ phases display absorption peaks centered around $\omega/v \simeq 0.10$--$0.15$. The longitudinal conductivities for the two Chern phases, \(C=+1\) and
\(C=-1\), are nearly overlapping. Therefore, although these phases have
opposite topological chiralities, their absorption spectra are almost indistinguishable. The magenta curve represents the longitudinal conductivity in the trivial phase, revealing a spin-dependent interband absorption response. The trivial phase does not exhibit a sharp dominant absorption peak like spin Chern and Chern-insulating phases; rather, at low photon energy, the conductivity rises and then shows a broad plateau. This pattern indicates that the optical transitions from the spin-up and spin-down sectors are allowed over partially separated but overlapping frequency windows. The weak low-energy pattern is associated with the smaller interband transition energy, while the broader high-energy response provides transitions that are involved in a larger spin-sector gap. Consequently, the trivial phase exhibits finite optical absorption, but its spectral weight is distributed more smoothly rather than appearing as two well-resolved peaks. Figure~\ref{Conductivity}(b) displays the imaginary part of the longitudinal conductivity. In the altermagnetic-spin-Chern phase, a strong dip near $\omega/\nu\simeq0.35$--$0.45$ can be seen, which is the dispersive counterpart of the sharp absorption peak in  $\mathrm{Re}\,\sigma_{xx}$. $C=+1$ and $C=-1$ phases exhibit nearly overlapping spectra because 
$\sigma_{xx}$ is insensitive to the sign of the Chern number. In contrast, the trivial phase displays a weaker, shifted feature arising from split spin-dependent transition energies.

Figures~\ref{Conductivity}(c) and (d) show the real and imaginary parts of the Hall
conductivity, respectively. The altermagnetic-spin-Chern phase exhibits a weak finite Hall response in the vicinity of the absorption threshold, while the two Chern phases show nearly opposite and much smaller Hall features. The trivial phase displays the largest Hall-resonant feature,  but its positive and negative Berry-curvature contributions compensate in the Brillouin zone, resulting in a zero net Chern number.  In the altermagnetic-spin-Chern phase, the black curve remains small and shows merely a weak dispersive feature near the absorption threshold, reflecting that the opposite spin sectors largely compensate for each other in the Hall response.  The two Chern phases show small, opposite low-energy features, consistent with their opposing Berry-curvature chiralities. In contrast, the trivial phase displays a much stronger spin-dependent resonance at higher energy, around $\omega/\nu\simeq 1.1$--$1.3$, which reflects the separation of its spin-resolved transition channels. Nevertheless, the integrated Berry curvature cancels, resulting in a net trivial phase with $C=0$.

\begin{figure*}[ht!]
	\centering	
	\includegraphics[width=1\linewidth]{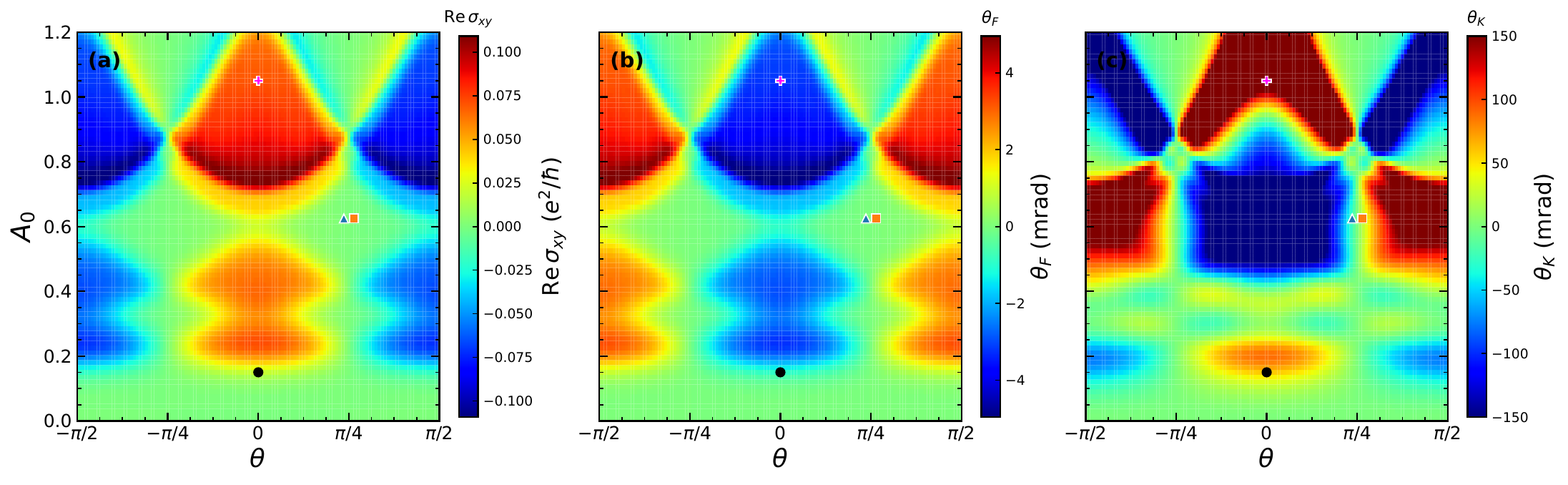}
\caption{(a) Real part of $\sigma_{xy}(\omega)$, (b) Faraday and (c) Kerr rotations  vs. $A_{0}$ and $\theta$.}
\label{contourcond}
\end{figure*}

Figure~\ref{contourcond}(a) demonstrates that the real part of $\sigma_{xy}$ is strongly anisotropic in the ($(A_0,\theta)$) plane. The opposite signs of the optical Hall response are shown in the red and blue regions,  reflecting Berry-curvature-weighted optical transitions of opposite chirality. In the altermagnetic-spin-Chern region at low-$A_0$, the two spin sectors largely compensate each other, so the net optical Hall conductivity remains relatively weak, as shown by the black point. In the Chern-insulating regions, marked by the triangular and square markers, this compensation is reduced. This is because one spin sector dominates the Hall response, and as a result, a finite $\mathrm{Re}\sigma_{xy}$ appears. Near the phase-boundary regions, where the Floquet-renormalized band gap approaches the probe photon energy, this response becomes noticeable. The trivial region is represented by the magenta diamond, showing a finite optical Hall response because interband transitions at finite-frequency remain active even when the dc topological Chern number is zero. Therefore, the enhancement of $\mathrm{Re}\sigma_{xy}$ should be interpreted as a combined effect of Berry curvature and resonance with the driven band structure.

\begin{figure}[ht!]
    \centering
    \includegraphics[width=1.00\linewidth]{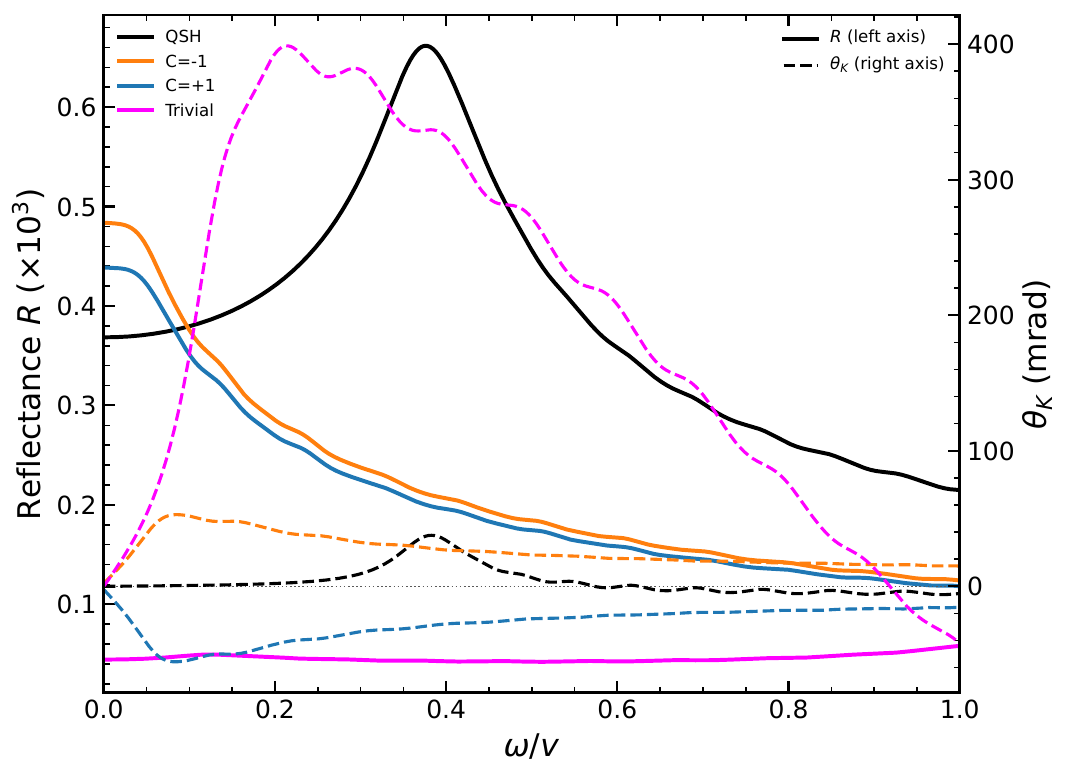}
    \caption{Frequency-dependent reflectance $\mathcal{R}$ (solid curves, left axis) and Kerr rotation $\theta_K$ (dashed curves, right axis) for representative points in the spin Chern, $C=-1$, $C=+1$, and trivial phases. The reflectance is displayed in units of $10^{-3}$ and the Kerr rotation in mrad. The $C=-1$ and $C=+1$ phases retain comparable reflected intensities but exhibit Kerr rotations of opposite sign, consistent with their opposite optical Hall chiralities. The trivial phase develops the largest resonant Kerr angle while its reflectance remains comparatively weak, demonstrating that a large Kerr rotation alone is not a sufficient indicator of nontrivial topology and must be interpreted together with $\mathcal{R}$ and the Kerr ellipticity.}
    \label{fig:representative_reflectance_kerr}
\end{figure}
Figure~\ref{contourcond}(b) presents the Faraday rotation angle ($\theta_F$) under the same conditions. It closely appears in the regions where the optical Hall response is strong. In the spin-Chern phase, the Faraday rotation is relatively small because the opposite spin-sector Hall responses cancel each others effect. In the Chern-insulating regime, this compensation is broken, producing a clear finite Faraday signal, with positive or negative rotation depending on the sign of $\mathrm{Re}\sigma_{xy}$. In the resonant regions where the fixed photon energy matches the Floquet-renormalized interband transition energy, the Faraday response appears at a large value. Although the trivial phase does yield a finite Faraday rotation, this signal is predominantly attributable to interband optical transitions at finite frequencies, as opposed to originating from a nontrivial topological Chern number.

\begin{figure*}[ht!]
	\centering	
	\includegraphics[width=1\linewidth]{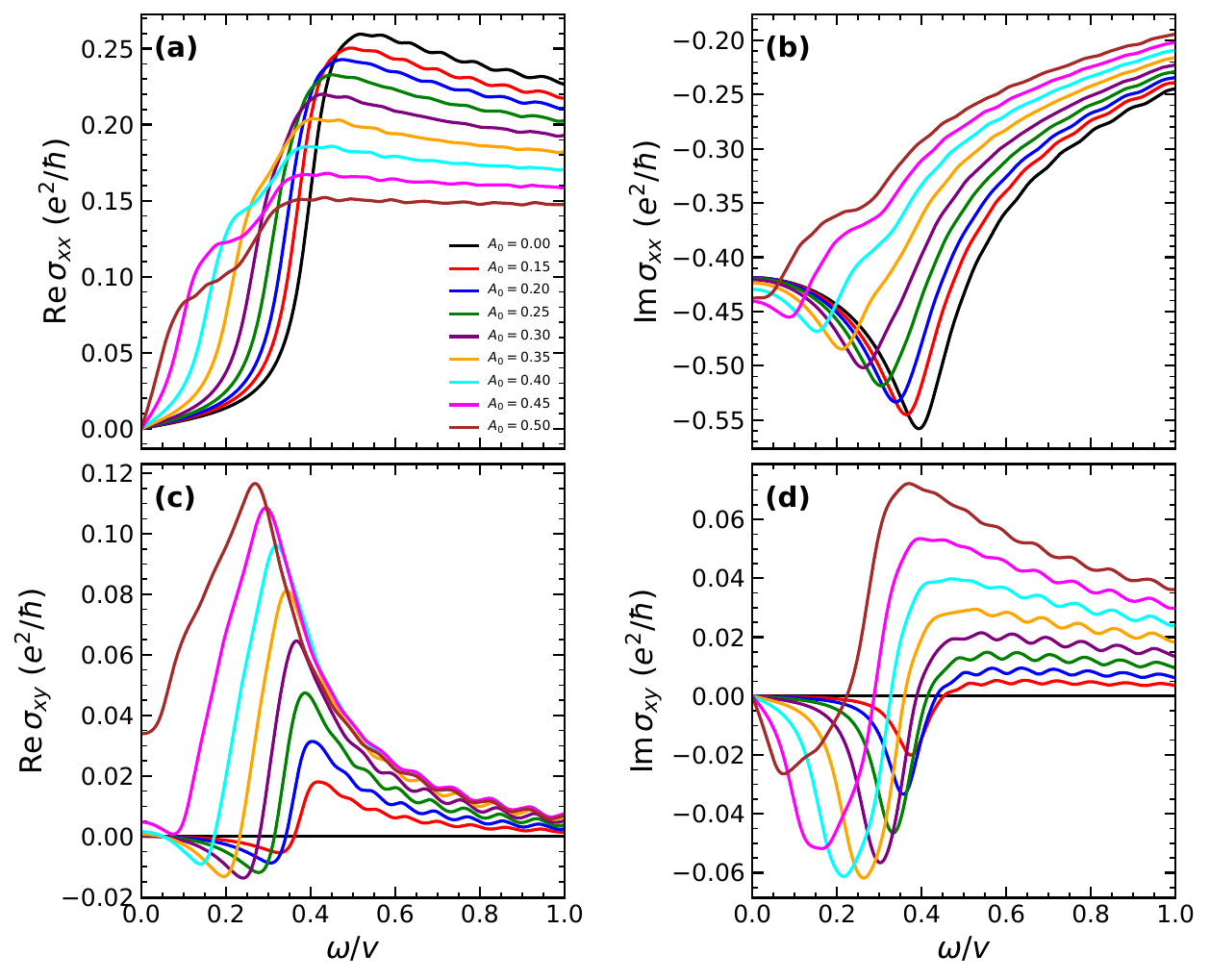}
\caption{(a) Real and (b) imaginary parts of $\sigma_{xx}$ vs. $\omega/v$ for different driving amplitudes $A_0$. (c) Real and (d) imaginary parts of $\sigma_{xy}$ vs. $\omega/v$ for different driving amplitudes $A_0$.}
\label{Cond1}
\end{figure*}
\begin{figure*}[ht!]
	\centering	
	\includegraphics[width=1\linewidth]{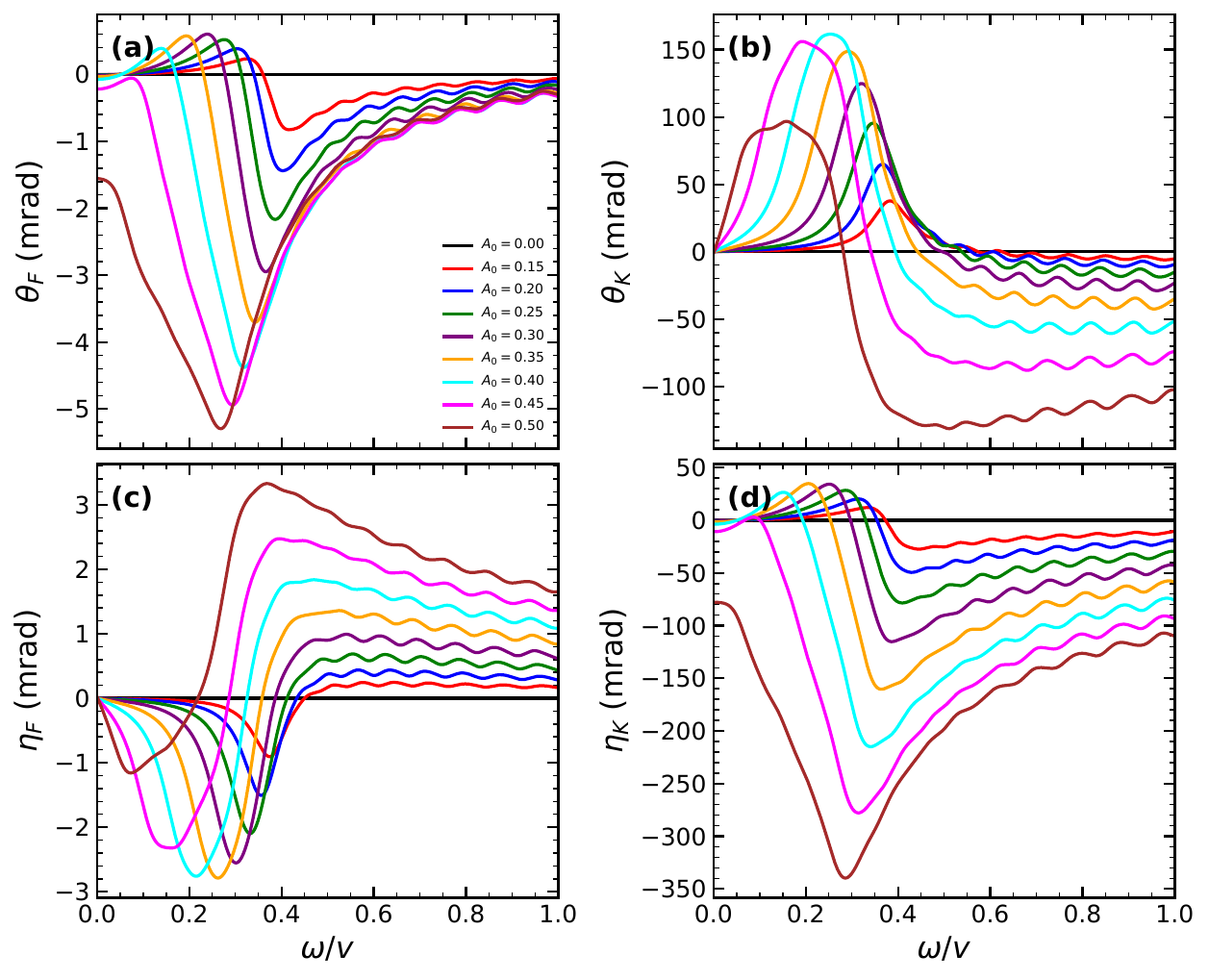}
\caption{(a) Faraday and (b) Kerr rotations vs. $\omega/v$ for different driving amplitudes $A_0$. (c) Faraday ellipticity and (d) Kerr ellipticity vs. $\omega/v$ for different driving amplitudes $A_0$.}
\label{FR2}
\end{figure*}
The Kerr rotation angle $\theta_K$ is shown in Figure~\ref{contourcond}(c). The comparison of Kerr rotation with the Faraday rotation shows that, in selected regions of the $(A_0,\theta)$ plane, the Kerr response is considerably stronger, reaching values on the order of $10^2$ mrad.  The Kerr rotation is comparatively weaker in the altermagnetic-spin-Chern region because of spin-sector compensation. In the Chern-insulating regime, particularly near resonant interband transitions, a large Kerr signal is produced due to the imbalance between spin-resolved Hall responses. In the trivial region, a considerable Kerr rotation can occur even if the dc topology is trivial because the reflected polarization remains highly sensitive to finite-frequency optical transitions and to the magnitude of the reflected field. The sign reversal of $\theta_K$ corresponds directly to a reversal of the effective optical Hall response. This confirms that Kerr rotation provides a complementary and highly sensitive probe of Floquet-engineered altermagnetic topology.

Figure~\ref{fig:representative_reflectance_kerr} provides the reflected-intensity check required for a physically meaningful interpretation of the Kerr spectra. The spin-Chern phase exhibits a broad reflectance maximum near $\omega/v\simeq0.4$, whereas its Kerr rotation remains comparatively moderate because the opposite spin-sector Hall responses largely compensate. The two Chern-insulating phases have similar reflectance envelopes over most of the plotted frequency interval, but their Kerr rotations have opposite signs. This behavior shows that the sign reversal between the $C=-1$ and $C=+1$ phases originates from the reversal of the optical Hall chirality rather than from a large difference in reflected intensity. Most importantly, the trivial phase produces a resonantly enhanced Kerr rotation approaching $4\times10^{2}$ mrad at low-to-intermediate frequency, while the corresponding reflectance is substantially smaller than in the spin Chern and Chern phases. The large trivial-phase Kerr angle therefore arises from the rapid phase variation of the weak reflected circular components near a finite-frequency resonance and should not be identified with a nonzero Chern number.

Figures~\ref{Cond1}(a)--(d) show the real and imaginary parts of the optical conductivity for different values of
the driving field amplitude $A_0$ at a fixed polarization angle $\theta=0$. Figure~\ref{Cond1}(a) displays the real part of $\mathrm{Re}\,\sigma_{xx}$ versus photon frequency. At low photon frequencies, the excitation energies of the spin-up and spin-down optical transitions are almost degenerate and thus dominate the optical response, producing a single strong peak. The absorption peak (or jump) is followed by a broad, plateau-like structure. By systematically increasing the strength of $A_0$, it can be seen that the optical absorption peak positions shift to the left. For large values of the optical field, each absorption peak splits into two. This splitting arises because the bandgaps of the spin-up and spin-down sectors are no longer degenerate under strong optical excitation. In Fig~\ref{Cond1}(b), we have displayed the imaginary part of the longitudinal conductivity. A negative sign over most of the frequency range signifies an inductive optical response in the driven system. The pronounced dips in $\mathrm{Im}\,\sigma_{xx}$ align spectrally with the maxima in $\mathrm{Re}\,\sigma_{xx}$. Figure~\ref{Cond1}(c) presents the real part of the Hall conductivity as a function of photon frequency. At low frequencies, the finite value of $\mathrm{Re}\,\sigma_{xy}$ indicates a nonzero anomalous Hall response. As the driving amplitude is varied,  both the position and magnitude of the Hall peaks change substantially. These peaks arise  when the photon energy coincides with Berry-curvature-active interband transitions. The imaginary part of the Hall conductivity is plotted in Fig~\ref{Cond1}(d).

Figure~\ref{FR2} displays the frequency dependence of the Faraday rotation $\theta_F$, Kerr rotation $\theta_K$, Faraday ellipticity $\eta_F$, and Kerr ellipticity $\eta_K$ for different values of the Floquet driving amplitude $A_0$. The spectra, presented as functions of the dimensionless photon frequency $\omega/v$, clearly demonstrate the strong tunability of the magneto-optical response by the external periodic drive. In Figure~\ref{FR2}(a), the Faraday rotation angle ($\theta_F$) is shown as a function of the normalized photon energy ($\omega/v$) for different Floquet driving amplitudes ($A_0$). In the absence of Floquet driving, $A_0=0$, the Faraday rotation vanishes, which is consistent with the absence of a net optical Hall response in the undriven case. Once the Floquet drive is applied, $\theta_F$ becomes finite and develops a clear dispersive structure. For small and intermediate $A_0$, the spectra first show a weak positive feature at low photon energy, followed by a stronger negative dip in the range $\omega/v\approx 0.25$--$0.45$. As $A_0$ increases, the position and magnitude of this dip change systematically, indicating that the Floquet drive renormalizes the interband transition energies and redistributes the Berry curvature in momentum space. The largest negative Faraday rotation appears for the stronger driving amplitudes, especially around $A_0=0.45$--$0.50$, where $\theta_F$ reaches values close to $-0.8$ mrad. At higher photon energies, $\theta_F$ gradually approaches smaller negative values, showing that the resonant Hall response weakens away from the dominant interband-transition region.

The Kerr rotation angle $\theta_K$ is represented in Figure~\ref{FR2}(b) under the same conditions. Unlike the Faraday rotation, the Kerr response is much larger, reaching values on the order of $10^2$ mrad. The $\theta_K$ reveals noticeable positive peaks at low-to-intermediate photon energies when the driving amplitudes are weak to moderate. As $A_0$ increases, these peaks shift, reflecting the Floquet modification of the optical transition spectrum. At higher values of $A_0$, specifically at $A_0=0.45$ and $A_0=0.50$, the Kerr rotation changes sign after the positive peak and evolves to strongly negative for a wide frequency range. This sign reversal shows a change in the effective phase difference between the two reflected circular polarization channels, and it is directly linked to the driven optical Hall response. Hence, Fig.~\ref{FR2}(b) illustrates that the reflected polarization is much more sensitive to Floquet driving than the transmitted polarization.

Figs.~\ref{FR2}(c) and~\ref{FR2}(d) show the corresponding ellipticities. The Faraday ellipticity ($\eta_F$), which measures the difference in absorption between the two circularly polarized transmitted components, is demonstrated in  Figure~\ref{FR2}(c). Therefore, a finite $\eta_F$ shows circular dichroism in transmission. Like $\theta_F$, the Faraday ellipticity approaches zero for $A_0=0$, evolving to finite values when the drive is applied. 
\begin{figure*}[ht!]
	\centering	
\includegraphics[width=1\linewidth]{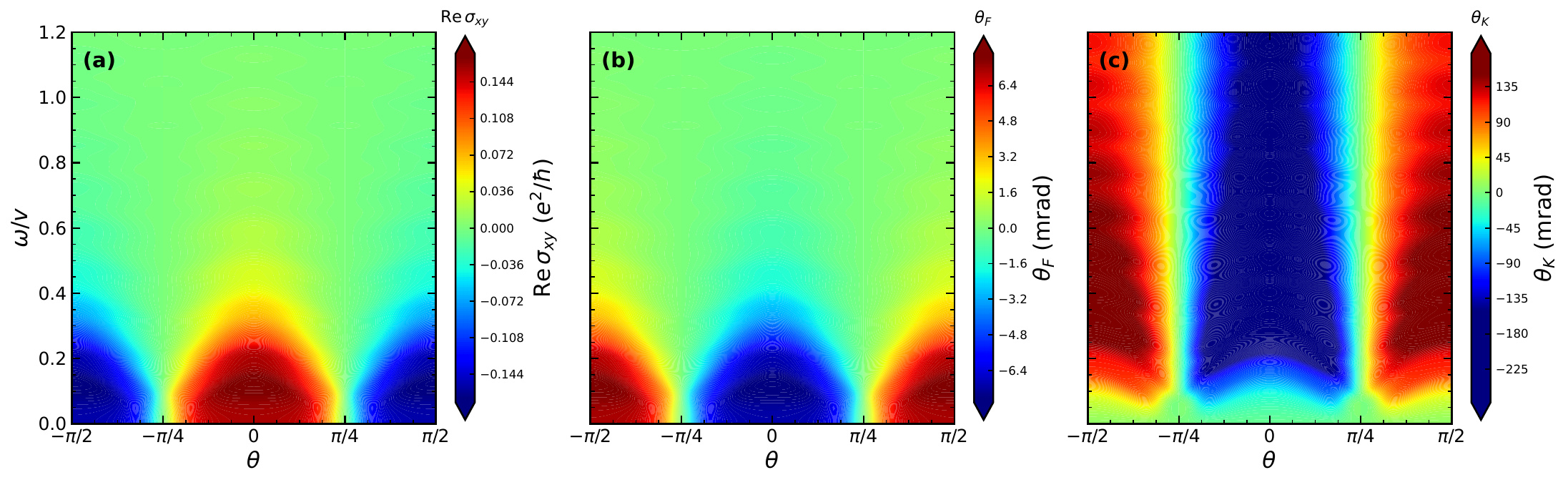}
\caption{(a) Real part of $\sigma_{xy}(\omega)$, (b) Faraday rotation angle and (c) Kerr rotation angle as a function of incident polarization angle and normalized incident photon frequency.}
\label{FR3}
\end{figure*}
At lower values of $A_0$, $\eta_F$ develops negative minima at low-to-intermediate photon energies. As $A_0$ increases, the negative behavior shifts and changes into a positive response at higher photon energies. For the largest amplitudes, particularly $A_0=0.45$ and $A_0=0.50$, $\eta_F$ becomes strongly positive, reaching values of approximately $0.4$-$0.5$ mrad over a wide range after the low-energy dip. This behavior shows that the phase difference between the circular components and their relative absorption is controlled by the Floquet drive. The comparison with $\eta_F$ displays that the Kerr ellipticity is much larger and remains mainly negative for all nonzero driving amplitudes. The magnitude of $\eta_K$ increases strongly with $A_0$, and the deepest negative minima occur in the same low-to-intermediate photon-energy window where the Kerr rotation changes rapidly. At high driving amplitudes, especially $A_0=0.45$ and $A_0=0.50$, $\eta_K$ reaches several hundred mrad in magnitude. This strong response originates from the combined effects of the optical Hall conductivity, longitudinal dissipative conductivity, and the sensitivity of the reflected field to small changes in the complex reflection coefficients.

In Figure~\ref{FR3}, the angular and frequency dependence of the real part of the optical Hall conductivity and the associated magneto-optical rotations as functions of the driving-field orientation $\theta$ and the normalized probe frequency $\omega/v$ are shown. Fig.~\ref{FR3}(a) shows that $\mathrm{Re}\sigma_{xy}$ exhibits a marked low-frequency structure with a strong positive lobe centered around $\theta\approx 0$ and negative lobes near $\theta\approx \pm \pi/2$. This clearly shows that the optical Hall response is highly anisotropic with respect to the direction of polarization of the driving field. In the low-energy regime, particularly for $\omega/v\lesssim 0.4$ the largest magnitude of $\mathrm{Re}\,\sigma_{xy}$ is seen. It is the point where the probe frequency is close to the Floquet-renormalized transition energies between bands. As the probe frequency increases, the Hall response quickly weakens and becomes nearly negligible in most of the $(\theta,\omega/v)$ plane. The sign reversal of $\mathrm{Re}\,\sigma_{xy}$ between the central and edge regions shows the redistribution of Berry curvature in the Floquet bands together with the change in the dominant optical transitions between bands.

Figure~\ref{FR3}(b) reflects the Faraday rotation angle $\theta_F$. The overall structure of $\theta_F$ nearly follows that of the optical Hall conductivity but with the opposite sign pattern; $\theta_F$ is predominantly negative around $\theta\approx 0$ and positive near $\theta\approx \pm \pi/2$. This behavior is consistent with the fact that the Faraday response is controlled by the phase difference between the two circularly polarized transmission channels $\sigma_{\pm}=\sigma_{xx}\pm i\sigma_{xy}$. Consequently, the strongest Faraday rotation appears in the same low-frequency region where Hall conductivity is enhanced. However, at higher $\omega/v$, the Faraday angle becomes very small, showing that the magneto-optical transmission response is concentrated mainly in the low-energy Floquet-renormalized optical window.

Fig.~\ref{FR3}(c) shows that the Kerr rotation angle $\theta_K$ has a much larger magnitude than $\theta_F$, reaching values on the order of $10^2$ mrad. In contrast to the Faraday map, the Kerr response is characterized by a broad negative region centered on $\theta\approx 0$ and positive regions near $\theta\approx \pm \pi/2$. This pattern is preserved over a wide frequency range, even though the largest magnitude is again found at low and intermediate frequencies. The enhanced Kerr response arises from the greater sensitivity of the reflected field to the complex optical Hall conductivity and to the impedance mismatch at the interface. Therefore, the sign reversal of $\theta_K$ across the $(\theta,\omega/v)$ plane reflects the Floquet-induced redistribution of Berry curvature that is amplified by the reflection geometry. From an experimental standpoint, the polarization-driven sign reversals of $\theta_F$ and $\theta_K$
 provide more robust phase markers than their peak magnitudes, which are sensitive to broadening effects and the surrounding electromagnetic environment. A polarization resolved pump--probe measurement at a fixed probe frequency could thus delineate the light-induced phase boundaries without the need for electrical contacts.

\section{Conclusions}
In this work, we have shown that linearly polarized light can be used to control the topological phases of a $d$-wave altermagnet. By varying the Floquet amplitude $A_0$ and the polarization angle $\theta$, the system exhibits transitions among spin Chern, trivial, and Chern insulating phases with $C=\pm1$. Physically, it is valid that linearly polarized light does not act through optical helicity. Instead, it produces polarization-dependent Floquet anisotropy that breaks the crystalline-time-reversal connection between the two altermagnetic spin sectors. This anisotropy results in spin-selective gap closings and reopenings, reverses the Berry-curvature chirality, changes the sign of the Hall conductivity, and generates tunable Faraday and Kerr responses. The largest magneto-optical signals occur near interband resonances and topological phase boundaries, where the optical Hall response is strongly enhanced. Therefore, Floquet $d$-wave altermagnets are promising candidates for switchable topological optics, reconfigurable Hall transport, and magneto-optical device applications without relying on conventional ferromagnetism. This chain of observables suggests a pump-probe strategy in which a strong linearly polarized pump prepares the Floquet phase and a weak polarization-resolved probe identifies it in real time.

%We acknowledge the financial support from the uni
%research Grant No. xyz.

%\section*{AUTHOR DECLARATIONS}
%The authors have no conflicts to disclose.

\renewcommand{\bibname}{References}
\bibliographystyle{apsrev4-2}
\bibliography{References}
\end{document}